\documentclass[journal=nanoletters,manuscript=article]{achemso}

\usepackage[version=3]{mhchem} % Formula subscripts using \ce{}
\usepackage{graphicx}
\usepackage{tabularx}
\usepackage{adjustbox}
\usepackage{threeparttable}
\usepackage{lipsum}
\usepackage{revquantum}
\usepackage{hyperref}
\usepackage{xcolor}
\usepackage{stackengine}

\usepackage{newfloat}
\DeclareFloatingEnvironment[name={Supplementary Figure}]{suppfigure}

\usepackage{mathtools}
\newtagform{supplementary}[S.]()
\counterwithin*{equation}{section}
\author{Lakshminarayan Sharma}
\author{Carlos Rodriguez-Fernandez}
\author{ Humeyra Caglayan}
\affiliation{Faculty of Engineering and Natural Sciences, Tampere University, Tampere 33880 Finland}
\email{humeyra.caglayan@tuni.fi}
\title[An \textsf{achemso} demo]
  {Fractional Dimensional Approach to Dielectric Tuning Effects on Excitonic Parameters in 2D semiconductor materials}

\begin{document}


\begin{tocentry}
\includegraphics[width=81mm]{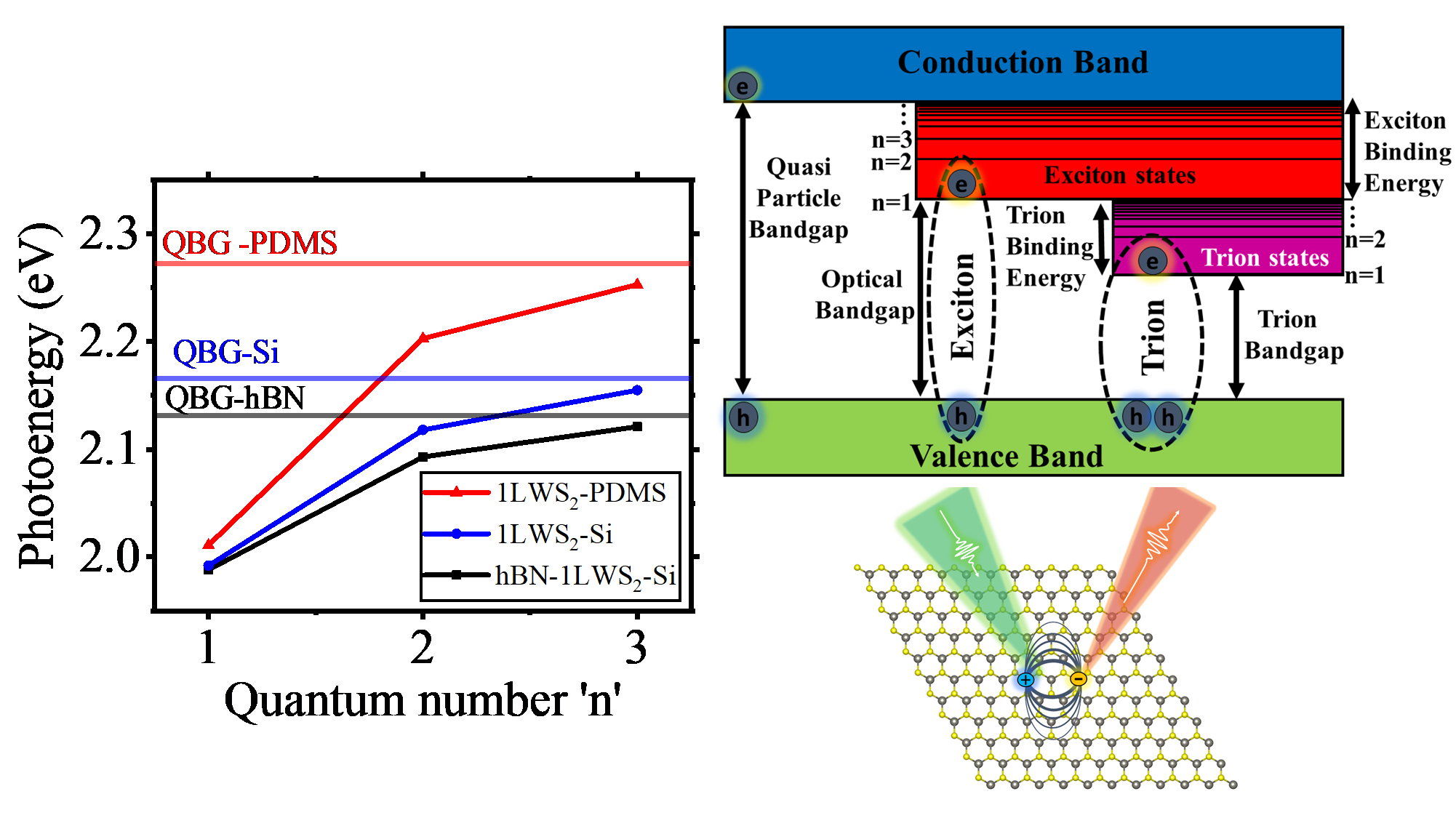}
\end{tocentry}

%TC:ignore
\begin{abstract}

%Our study showcases the remarkable potential of the fractional dimensional approach in unraveling the exciton parameters within the prototypical atomically thin semiconductor, a monolayer of WS$_2$. We demonstrated its effectiveness in accurately gauging the exciton binding energy and quasiparticle bandgap of the WS$_2$ monolayer across diverse dielectric environments. By altering the dielectric environment from 1.52 to 8.1, we achieved a tuning of the quasiparticle bandgap and binding energy by 141 meV and 188 meV, respectively. Crucially, the fractional dimension deduced from the excitonic Rydberg series consistently hovers around 2.8 for the WS$_2$ monolayer, regardless of the dielectric milieu. This discovery underscores the method's agility and reliability in swiftly determining exciton binding energy in 2D semiconductors, transcending the constraints of specific dielectric environments. In essence, our research offers a powerful and versatile tool for investigating exciton binding energy in 2D semiconductors with profound implications for diverse applications.

We demonstrated the potential of the fractional dimensional approach to understand exciton parameters in the exemplary atomically thin semiconductor material, a monolayer of WS$_2$. This approach has proved to be successful in finding the exciton binding energy and quasiparticle bandgap for the WS$_2$ monolayer in varying dielectric environments. A tuning of the quasiparticle bandgap and binding energy by 141 meV and 188 meV, respectively, has been achieved by varying the dielectric of the environment from 1.52 to 8.1. The approach is justified by comparing the changes in the binding energy with the computational results from the Quantum Electrostatic Heterostructures model. The fractional dimension found through the excitonic Rydberg series is close to 2.8 for WS$_2$ monolayer in all different dielectric surroundings. Thus, this approach provides a rapid and robust method for determining the binding energy of excitons in 2D semiconductors independent of the particular dielectric environment.

%Hence, this approach offers a swift means of determining the binding energy of excitons in 2D semiconductors, irrespective of the specific dielectric environment.
 
\end{abstract}
%TC:endignore

\section{Introduction}
Two-dimensional transition metal dichalcogenides (2D-TMDCs) exhibit a plethora of remarkable and intriguing properties. Notable among these is the transition from an indirect to a direct bandgap in monolayers, a substantial magnitude of spin‐orbit splitting, the phenomenon of valley polarization, and the emission of single photons  \cite{Mak2010, Mak2012, Splendiani2010, Radisavljevic2011, Gao2023, He2015}.  The presence of tightly bound electron‐hole pairs, known as excitons, is central to the optical characteristics of these atomically thin materials.
The excitons found in these materials possess considerable binding energies. This notable attribute arises from the combination of reduced dielectric screening in two dimensions and the quantum confinement effects inherent to such systems \cite{Wang2018}. One particularly fascinating aspect is the pronounced sensitivity of the exciton binding energy to the spatial dielectric environment \cite{Raja2017, Raja2019}. The electrical manipulation of these excitonic states enables the tuning of light‐matter response \cite{Chernikov2015, Li2023, Gherabli2023, Tsai2023, Liu2020, Tagarelli2023, Lee2017, Ross2014, Cheng2014, Withers2015, Seyler2015, He2015, Wang2018} and creation of atomically thin devices such as transistors \cite{Radisavljevic2011, Yin2011}, photodetectors \cite{LopezSanchez2013, Zhang2016}, p-n diode \cite{Baugher2014, Pospischil2014, Lee2014}, LEDs \cite{Pospischil2014, Ross2014,Frisenda2018} and solar cells \cite{Pospischil2014, Furchi2014,Frisenda2018}.

%Furthermore, thecapacity for electrical manipulation of these excitonic states enables the tuning and customization of the light‐matter response \cite{Chernikov2015, Li2023, Gherabli2023, Tsai2023, Liu2020, Tagarelli2023, Lee2017, Ross2014, Cheng2014, Withers2015, Seyler2015, He2015, Wang2018}. 

While the concept of dielectric engineering in 2D materials is well-established, the intricate relationship between the dielectric environment and specific excitonic parameters, such as the exciton binding energy (BE) and quasiparticle bandgap (QBG), has remained a subject of intense investigation \cite{Raja2017, Chernikov2014, Hill2015, Raja2019, Rigosi2016, Stier2016a, Hsu2019a, Shi2022, Stier2018, Aslan2021, Wang2020}. The quantitative knowledge of both BE and QBG is necessary for optimizing 2D materials for various electronic and photonics applications. Addressing this challenge conventionally involves employing integer-dimensional models, but these models frequently prove inadequate in capturing the unique attributes of 2D TMDCs. These characteristics encompass the monolayer's finite thickness in the out-of-plane direction and the abrupt potential changes that manifest at the interfaces within the monolayer. To overcome this limitation, new approaches based on integrating a 
quasi-2D Coulomb potential (such as the Ohno Potential \cite{Meckbach2018}) or truncating the Coulomb potential \cite{Qiu2016} have been proposed. However,  these all require extensive numerical solutions. Yet, the fractional dimensional approach (FDA), which emerges as a powerful method, precisely accounts for the finite out-of-plane extension's influence. However,  these all require extensive numerical solutions. Yet, the fractional dimensional approach (FDA), which emerges as a powerful method, precisely accounts for the finite out-of-plane extension's influence by not restricting the system to an absolute 2 dimensions. 

%Our work takes into account the finite out-of-plane thickness by not restricting the system to an absolute 2 dimensions. An ingenious approach that allows the possibility to use fractional dimensions between 2-3 is being utilized.This work employs a fractional dimensional approach (FDA) to precisely account for the finite out-of-plane extension's influence. 

The FDA is a formalism based on the non-integer (Hausdorff) dimension - $\alpha$ in which the principal quantum number `n' remains a valid quantum number for the Coulomb Potential \cite{Stillinger1977, He1991, Christol1993}. The fractional dimension results from the exciton's constrained motion in an atomically thin material due to the exciton-lattice interaction \cite{He1991, Christol1993,Lefebvre1993}. Thus, the fractional dimension manifests the exciton dynamics \cite{He1991, He1990a}. Since its inception in 1990 \cite{He1990,He1990a, He1990b}, the FDA has been applied to the study of the excitons in quantum well \cite{Christol1993, Lefebvre1993, Thilagam2001, Mathieu1992, MatosAbiague1998, Thilagam1997}, quantum wires \cite{Christol1993, Lefebvre1993}, exciton-phonon interaction \cite{Thilagam1997}, biexcitons \cite{Oh1999} and exciton-exciton interaction \cite{Thilagam2001}. 

%In this study, we show, for the first time as of our knowledge, the experimental application of this method to the dielectric tuning of exciton BE and QBG in 2D TMDCs.

In contrast to the numerical approach for solving the Schr$\Ddot{o}$dinger equation in 2D dimension with Rytova-Keldysh potential, the FDA provides an analytical and systematic solution to the problem. %It provides a straightforward way to calculate the change in exciton binding energy and, thus, the QBG due to the effect of the external dielectric surroundings. 
This approach represents a lightweight analytical model with efficient and excellent tunability as the fractional dimension $\alpha$ inherently encompasses the anisotropy caused by the environment and the influence of the perturbative potential interacting with the exciton \cite{Lefebvre1993, Christol1993}. Along with this, when the first and second excited states of the exciton transition energies are known, the method enables the direct calculation of the BE \cite{Mathieu1992}. 

%In this work, a combined approach consisting of linear reflectance spectroscopy and FDA has been employed. We demonstrate an effective means to measure exciton BE and QBG and account for dielectric environment effects. This integrated approach offers a streamlined and straightforward avenue for determining exciton parameters. This provides a great insight into the tunable nature of the optical gap and QBG depending on the dielectric environment of the 2D TMDCs, which can be used to create new photonic and optoelectronic devices with tailorable light matter response. Along with this, our approach readily applies to diverse 2D materials, accommodating variations in the dielectric environment. Additionally, this suggested method enables finding the overall band type for atomically thin horizontal van der Waals heterostructure.  
%Enabling usefulness to find the overall band type for %atomically thin horizontal van der Waals heterostructure.  

In this study, we employ a combined approach of linear reflectance spectroscopy and the FDA to effectively measure exciton BE and QBG while considering dielectric environment effects. We show, for the first time as of our knowledge, the experimental application of this method to the dielectric tuning of exciton BE and QBG in 2D TMDCs. This integrated methodology provides a streamlined pathway for determining exciton parameters in 2D TMDCs. Our approach not only reveals the excitonic landscape in these materials but also sheds light on the tunable nature of the QBG, contingent on the dielectric environment of 2D TMDCs. %These insights hold promise for tailoring the light-matter response in photonic and optoelectronic devices. 
This method can be a universal tool for exploring excitonic properties in diverse 2D materials. Additionally, it uniquely allows for determining the overall band type in atomically thin horizontal van der Waals heterostructures, enhancing our understanding of their electronic structure.

\begin{figure} [ht] 
\centering\includegraphics[width=1.0\linewidth]{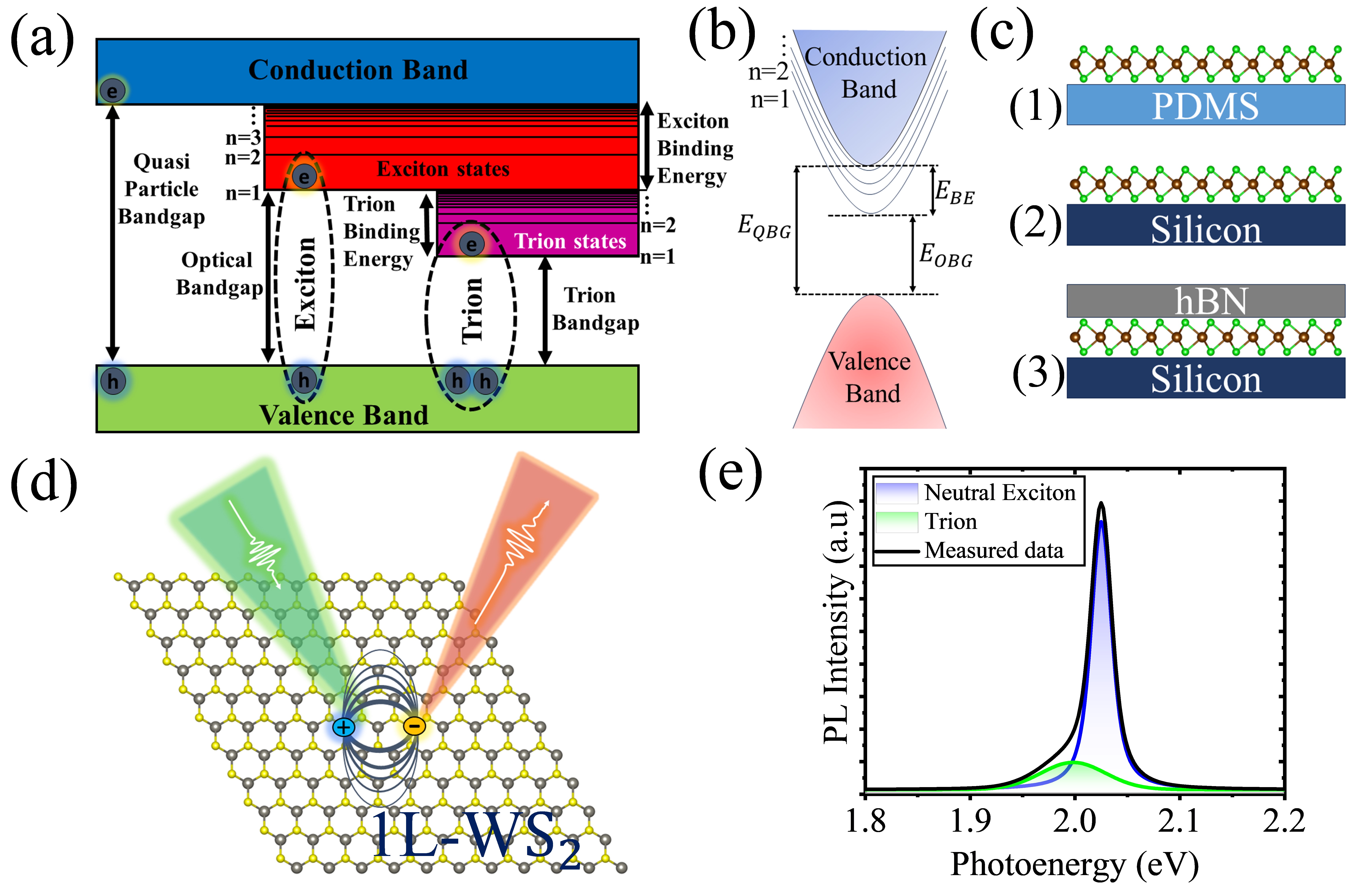}
\caption{(a) Illustration of the electronic bandgap and the exciton Rydberg series. Both the trion and exciton states are shown. Along with this, the key fundamental parameters, such as trion BE, optical bandgap (OBG), exciton BE, and the QBG, are represented. (b) Simplified schematic of (a). (c) Diagram of the various substrate structures studied in this work. (d) An illustration for the excitation of WS$_2$ monolayer by 2.330 eV (532 nm) photon and the excitonic emission at 2.025 eV (612 nm). (e) The PL spectrum for 1L-WS$_2$, the trion emission at 1.999 eV, is shown in green.}
\label{fig1}
\end{figure}
An illustration of the electronic bandgap and the exciton Rydberg states are shown in Figure \ref{fig1} (a) and (b). Due to quantum confinement and reduced dielectric screening, the Coulomb attraction between the charge carriers is strongly enhanced at the nanoscale regime. This creates a neutral two-body quasiparticle with energy lower than the quasiparticle (single particle) bandgap. The quantum states of the exciton are labeled by `n' throughout the manuscript. %The energy difference between the QBG and the exciton transition energy defines the BE for the respective exciton state `n'.
The term BE will correspond to the ground state (n=1) exciton binding energy throughout the paper. 

%The fundamental optical excitation in 2D semiconductors corresponds to the n=1 state of the exciton termed as the optical bandgap to distinguish it from the QBG. 

\section{Results and discussion}

In this work, we use the FDA method to investigate the dielectric tuning of excitonic parameters in monolayer (1L) WS${_2}$. We explore three distinct dielectric environments: (1) monolayer on a polydimethylsiloxane (PDMS) substrate, (2) monolayers directly on a silicon substrate, and (3) top hBN (13.28 nm) encapsulated on a silicon substrate, for (1) and (2) air is the top dielectric medium. A schematic representation of the various substrates used in this study is shown in Figure \ref{fig1} (c). To quantify the dielectric environment, we define the average dielectric constant of the environment surrounding the monolayer as $\kappa$ = $\epsilon_{top} + \epsilon_{bottom}$/2. For PDMS, Si and hBN-Si substrate, we have $\epsilon_{PDMS}$ = 2.04 \cite{Schneider2009}, $\epsilon_{Si}$ = 11.7 \cite{Dunlap1953}, $\epsilon_{hBN}$ = 4.5 \cite{Geick1966}, which gives $\kappa$ as 1.52, 6.35 and 8.1 respectively. These configurations allow us to systematically assess the influence of different dielectric conditions on excitonic properties in 2D TMDCs. The photoluminescence (PL) spectrum for 1L WS$_2$ at room temperature is shown in Figure \ref{fig1} (e), which was obtained by pumping at 532 nm. A schematic representation of the excitation of the 1L WS$_2$ at 2.330 eV (green) and its excitonic emission at 2.025 eV (red) is shown in Figure \ref{fig1} (d). In the collected PL spectrum, we found the presence of trion at 1.999 eV, represented by the green peak at the lower energy in Figure \ref{fig1} (e). From the difference between the exciton and trion emission peaks, we calculated the trion binding energy of 26 meV.   

\begin{figure}[ht]  
\centering\includegraphics[width=1.0\linewidth]{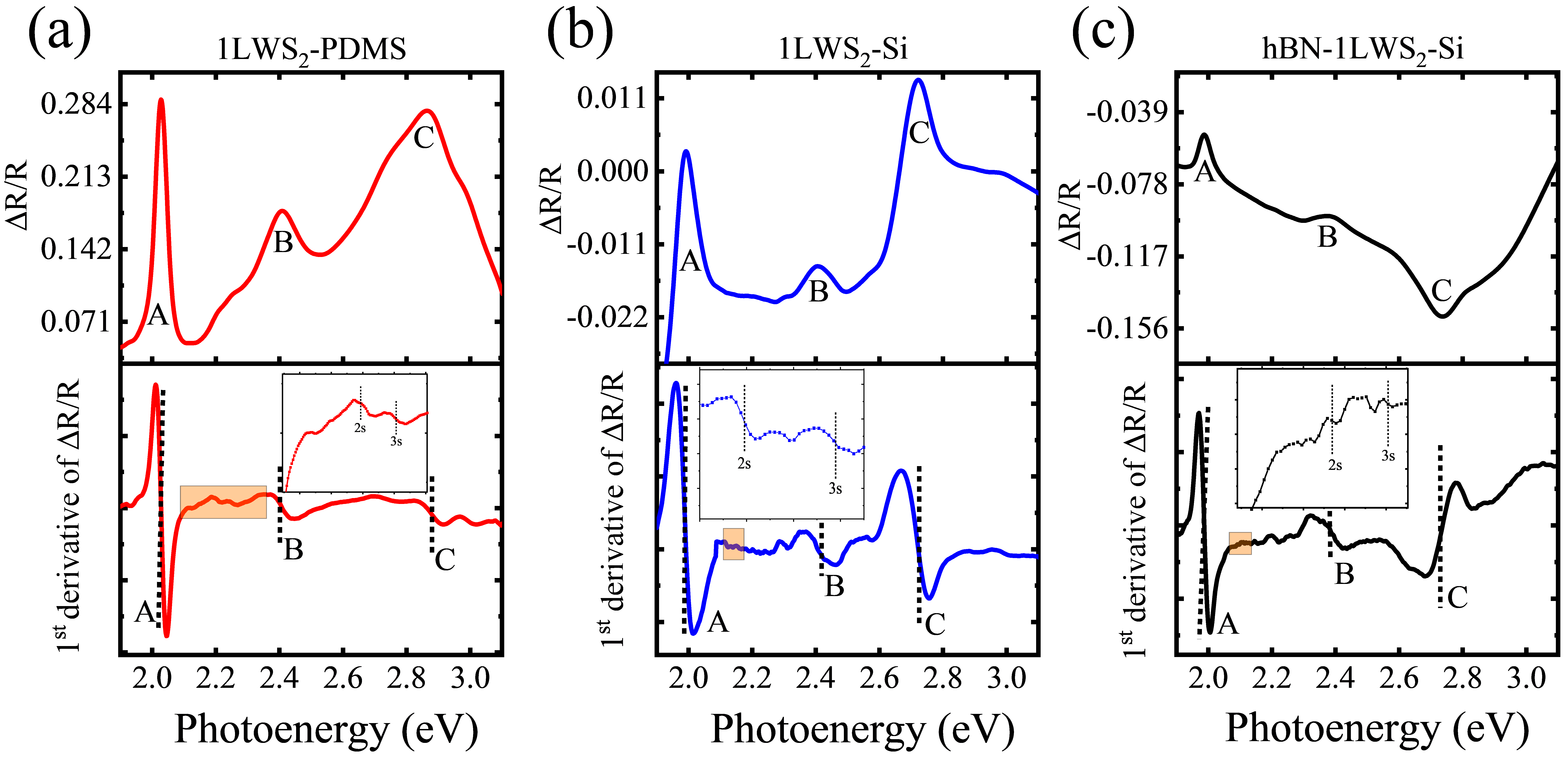}
\caption{Measured reflectance contrast ($\Delta R/R$) spectra and its first derivative for monolayer WS$_2$ on (a) PDMS, (b) Si and (c) top hBN encapsulation on Si substrate. The top panel plots label the A, B, and C exciton peaks. The position of the A, B, and C exciton, along with the A exciton higher energy states 2s and 3s are marked by dashed lines in the bottom panel.}
\label{fig2}
\end{figure}

In tungsten (W) based 2D TMDCs, the energy separation between the spin-orbit split A and B exciton is large enough to observe the A exciton Rydberg series \cite{Zhao2013, Chernikov2014}. This enables the determination of the exciton binding energy and the quasiparticle bandgap by fitting the experimentally found excitonic Rydberg series to the fractional dimensional model. To experimentally determine the excitonic Rydberg series, we carried out straightforward and effective reflectance contrast ($\Delta R/R$) spectroscopy, eliminating the need for complex methods like two-photon or magneto-optical characterization. The reflectance contrast (RC) is the normalized difference between the reflectance from the atomically thin sample and the substrate. $\Delta R/R = (R_{sample} - R_{substrate})/(R_{substrate})$, in the limit of an atomically thin sheet of materials, depends on the real part of the conductivity which is proportional to the absorption \cite{Li2018}. The RC spectrum for the three different samples of monolayer WS$_2$ is shown in Figure \ref{fig2} (a)-(c). The three major peaks in those spectra correspond to the primary exciton transition A, B, and C. The A exciton is the fundamental excitation corresponding to n = 1 states, occurring at the hexagonal Brillouin zone's K and K$^\prime$ valleys. The A exciton transition was at 2.011 eV, 1.992 eV, and 1.988 eV for 1L WS$_2$ on PDMS, Si, and top hBN encapsulation on Si substrate, respectively. This corresponds to a redshift of 23 meV for the A exciton peak when changing the dielectric constant of the environment from that of a PDMS to hBN-Si. This effect of the external dielectric environment on the monolayer is such that the reduction in binding energy adjusts the QBG renormalization. Thus, the optical band gap (1s state energy) remains relatively unchanged (23 meV shift). The B exciton peak originates from the spin-orbit split valence band transition at the K and  K$^\prime$ valleys of the hexagonal Brillouin zone. The separation between A and B exciton provides the value of spin-orbit splitting, which was found to be around 400 meV. The origin of C exciton is due to the singularities in the joint density of states (JDOS) known as van Hove singularities in the band diagram, which leads to band nesting regions where a small section or region of the band structure belonging to the conduction and valence band are parallel to one another \cite{Gillen2017, Carvalho2013, Mennel2020, Goswami2021}.

%In particular, it corresponds to the band nesting region in
%between along the $\Gamma‐\Lambda$ (midpoint between $\Gamma‐K$ points) line of the Brillouin zone \cite{Goswami2021}. 

Here, our primary focus is on the A exciton and its Rydberg series. To reveal the minute spectral features of the higher energy A exciton Rydberg states, we took the  1$^{st}$ derivative of the RC spectrum as shown in Figure \ref{fig2} (a)-(c) bottom panel. In these 1$^{st}$ derivative plots, at the position of A, B, and C exciton peaks, there is a clear change in the curvature, and these inflection points are indicated with dashed lines. Observing the regions between the A and B exciton inflection points, there are additional points of change in curvature, which correspond to higher energy states of the A exciton as highlighted in Figure \ref{fig2} (a)-(c) bottom panel. The excited state n = 2 was observed at 2.203 eV, 2.118 eV, and 2.093 eV on PDMS, Si, and top hBN encapsulation on the Si substrate, respectively. The excitonic Rydberg states (2s and 3s) for the three samples are shown in the inset of Figure \ref{fig2}(a)-(c) bottom panel, where the peak position is shown by the dashed lines representing the inflection point. In comparison to the energy shift of the 1s state, the 2s state has redshifted by a much larger amount, 110 meV, because of the renormalization of the QBG due to an increase of the dielectric constant of the surroundings from PDMS to hBN-Si. This shift of the 100s of meV is a strong indication of the effect of the dielectric environment on the BE and QBG. 

The BE is the difference between the QBG and the exciton Rydberg state energy (RSE): 
\begin{equation}\label{eq1}
E_n^{BE}= E_{QBG} - E_n^{RSE}
\end{equation} 
The eigenenergy solution for the non-integer dimensional space $\alpha$ is given by \cite{He1991} 
\begin{equation}\label{eq2}
E_n^{BE}=-\frac{\mu R_y}{ m_e \epsilon_{eff}^2 } \dfrac{1}{(n+ \dfrac{\alpha-3}{2})^2}
\end{equation} 
where $\mu$ the reduced mass of the exciton, R$_y$ is the Rydberg energy (13.6 eV), m$_e$ is the free electron mass, $\epsilon_{eff}$ is the effective dielectric constant seen by the exciton. For integer dimensions where $\alpha$ = 3 and 2, we get the known $\dfrac{1}{n^2}$ and $\dfrac{1}{(n-1/2)^2}$ dependence, respectively \cite{Mathieu1992}. We calculated the BE and QBG for the three samples by fitting the RSE for the A exciton, using Equations \ref{eq1} and \ref{eq2}. 

\begin{figure} [ht] 
\centering\includegraphics[width=1.0\linewidth]{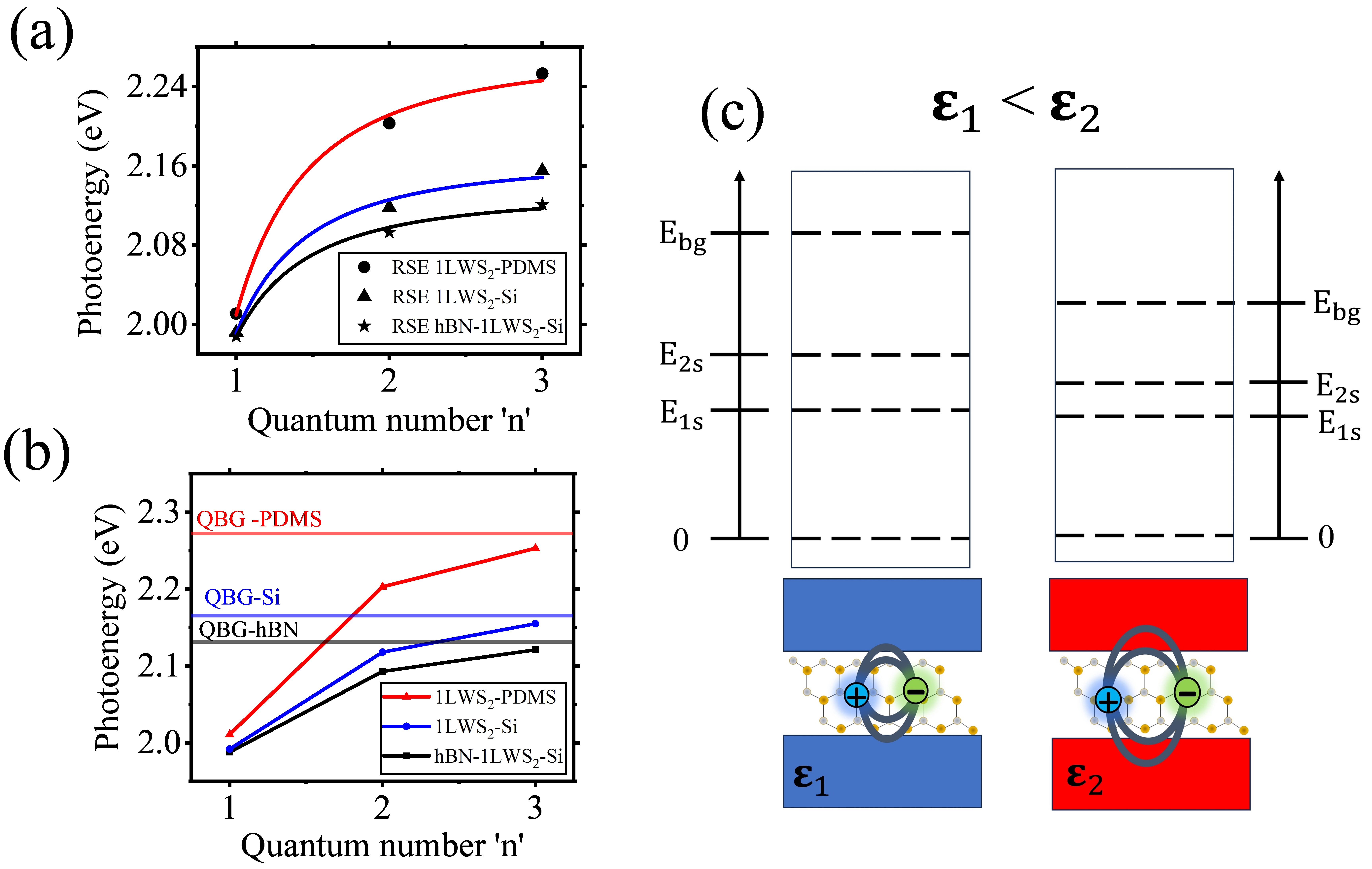}
\caption{ Experimentally measured exciton Rydberg state energies (RSE) for 1L WS$_2$ on the various substrates as a function of `n'. (a) The red, blue, and black lines represent the FDA fit to the RSE data points for PDMS, silicon, and top hBN encapsulation on a silicon substrate. (b) The horizontal red, blue, and black line represents the  QBG for PDMS, silicon, and top hBN encapsulation on a silicon substrate, respectively, as calculated from the FDA fit. 
(c) Schematic illustration of the effect of dielectric surroundings on the exciton spatial size and the RSE. The modification of the strength of the Coulomb interaction between electron-hole pair in two different dielectric surroundings with $\epsilon_1 < \epsilon_2$ leads to a decrease of the QBG and BE.   }
\label{fig3}
\end{figure}

By substituting Equation \ref{eq1} in Equation \ref{eq2}, we get the final form of the equation for the FDA fit. This final form contains four unknowns, $\mu, \alpha, \epsilon_{eff}$, and the QBG. For the fit, we used the exciton reduced mass of 0.16 m$_e$ for 1L WS$_2$ reported using magneto-optical spectroscopy \cite{Stier2016a}. To ascertain the value of $\alpha$, we computed the following ratio $\dfrac{E_{3s} - E_{1s}}{E_{2s} - E_{1s}}$, which only depends on the fractional dimension $\alpha$ (see supplementary Section 2). Thus, the two unknown E$_{QBG}$ and $\epsilon_{eff}$ were set as free parameters, and their values were obtained through the fit. The FDA fit to the measured excitonic Rydberg series for the three samples is shown in Figure \ref{fig3} (a). For 1L WS$_2$ on PDMS, Si and top hBN encapsulation on Si substrate, the $\epsilon_{eff}$ is obtained from the fit as 3.080 $\pm$0.75, 3.87 $\pm$ 1.13, 4.18 $\pm$ 1.06, respectively. The calculated QBG from FDA and the exciton RSE series are also shown in Figure \ref{fig3} (b). The details of the FDA for the three samples are summarized in Table \ref{Table1}. To understand the prominent effect of the dielectric surrounding on the exciton properties, we need to visualize the electric field distribution and strength between an electron and the hole that constitutes an exciton. A schematic diagram representing this is shown in Figure \ref{fig3} (c) for two different dielectric surroundings with $\epsilon_1 < \epsilon_2$. For a higher dielectric constant of the surrounding, there is a larger screening of the Coulomb interaction between the electron-hole pairs, which reduces the strength of the attractive potential and leads to an increase in the overall size of the exciton \cite{Stier2016}. This leads to a decrease in the QBG and BE magnitude. 

%The dielectric screening length is related to the E$^{BE}$ as follows \cite{NguyenTruong2022}: 
%\begin{equation}\label{eq21}
%r_0=\frac{5}{2 E_{\mathrm{b}}}\left[\sqrt{2}-\left(\frac{2 \epsilon_{eff}^2}{\mu} E_{\mathrm{b}}\right)^{1 / 4}\right]^2
%\end{equation}
%Using the $\epsilon_{eff}$ values as found from the FDA fit, we get $r_0$ equal to 38.14 \AA, 46.61 \AA, and 67.07 \AA, for the PDMS, silicon, and hBN on top samples, respectively. These values are consistent with the reported values for  $r_0$ = 32.99 \AA \cite{Olsen2016} and 37.89 \AA \cite{Donck2018}. Thus, we observe a decrease in the E$^{BE}$ with increasing  $r_0$ as established by theoretical works \cite{NguyenTruong2022}.   

\begin{table} [ht]
\centering
\begin{tabular}
{|l|l|l|l|l|}
\hline \text { Substrate } & \text { $\alpha$ } & \text { QBG (eV) } & \text { BE (meV) } & \text { $\kappa$ } \\
\hline \text { 1L WS$_2$-PDMS } & 2.86 & 2.272 $\pm$ 0.009  & 261 $\pm$ 9 & 1.52 \\
\hline \text { 1L  WS$_2$-Si } & 2.82 & 2.165  $\pm$ 0.008 & 173 $\pm$ 8 & 6.35 \\
\hline \text { hBN-1L  WS$_2$-Si } & 2.85 &  2.131 $\pm$ 0.005 & 143 $\pm$ 5 & 8.1 \\
\hline
\end{tabular}
\caption{FDA results for the three samples used in this study. The fractional dimension $\alpha$, QBG, B.E, and the average dielectric constant of the environment $\kappa$  are summarized.}
\label{Table1}
\end{table}

The bulk (3D) exciton BE (E$_{3D}$) is linked to the BE as calculated through the FDA by the following relationship\cite{Christol1993, Mathieu1992}: 
\begin{equation}\label{eq3}
\frac{E^{FDA}}{E^{3D} } 
\approx \frac{108}{ 101 } \dfrac{E_{2s}- E_{1s}} {E_{3D}} + \frac{20}{101} 
\end{equation}

Using the experimentally found E$_{3D}$ for WS$_2$ of 57 meV \cite{Beal1976} and the E$_{2s}$ - E$_{1s}$ value as found in this work, we determined the value of E$^{FDA}$ as 216 meV, 146 meV, and 123 meV for 1L WS$_2$ on PDMS, silicon, and hBN-Si substrates. Upon inspection, it becomes clear that these values are in accordance with the binding energy obtained from the FDA fit, as summarized in Table \ref{Table1}. We also applied Equation \ref{eq3} to find the E$^{FDA}$ for 1L WSe$_2$. For which, we utilized the established  E$_{3D}$ value of 55 meV \cite{Beal1976}, coupled with the energy difference between E$_{2s}$ - E$_{1s}$, which amounts to 130 meV \cite{Stier2018} and 131 meV \cite{Wang2020} from two distinct research studies. Our calculations yielded E$^{FDA}$ values of 149 meV and 150 meV, closely mirroring the experimentally observed values of 161 meV \cite{Stier2018} and 168 meV \cite{Wang2020} respectively, reported in the original studies.

The system's dimensionality can be measured by adopting a physical parameter dependent on the dimension as a valid criterion \cite{He1990}. Here, we have selected the RSE's binding energy ratio as such a criterion. The binding energy ratio for the 1s:2s:3s states for the PDMS, Si, hBN-Si substrates are $1:\dfrac{1}{3.78}:\dfrac{1}{13.73}; 1:\dfrac{1}{3.68}:\dfrac{1}{17.30}$; and 1:$\dfrac{1}{3.76}:\dfrac{1}{14.3}$, respectively. In contrast, when considering the 2D and 3D systems with a Coulomb potential (1/r), the binding energy ratios for the 1s:2s:3s states are $1:\dfrac{1}{9}:\dfrac{1}{25}$; and 1:$\dfrac{1}{4}:\dfrac{1}{9}$, respectively \cite{Goryca2019}. Notably, the dimensionality of monolayer WS$_2$ within the three unique dielectric environments consistently converges to a fractional dimension of approximately 2.8 and perfectly aligns with the exciton Rydberg series. Upon comparison, it also becomes apparent that the observed binding energy ratios for the three samples closely mirror the ratios anticipated in a three-dimensional (3D) model. This robustly corroborates our findings, reinforcing the dimensionality of approximately 2.8, as unveiled by our research. 

\begin{figure} [ht] 
\centering\includegraphics[width=1.0\linewidth]{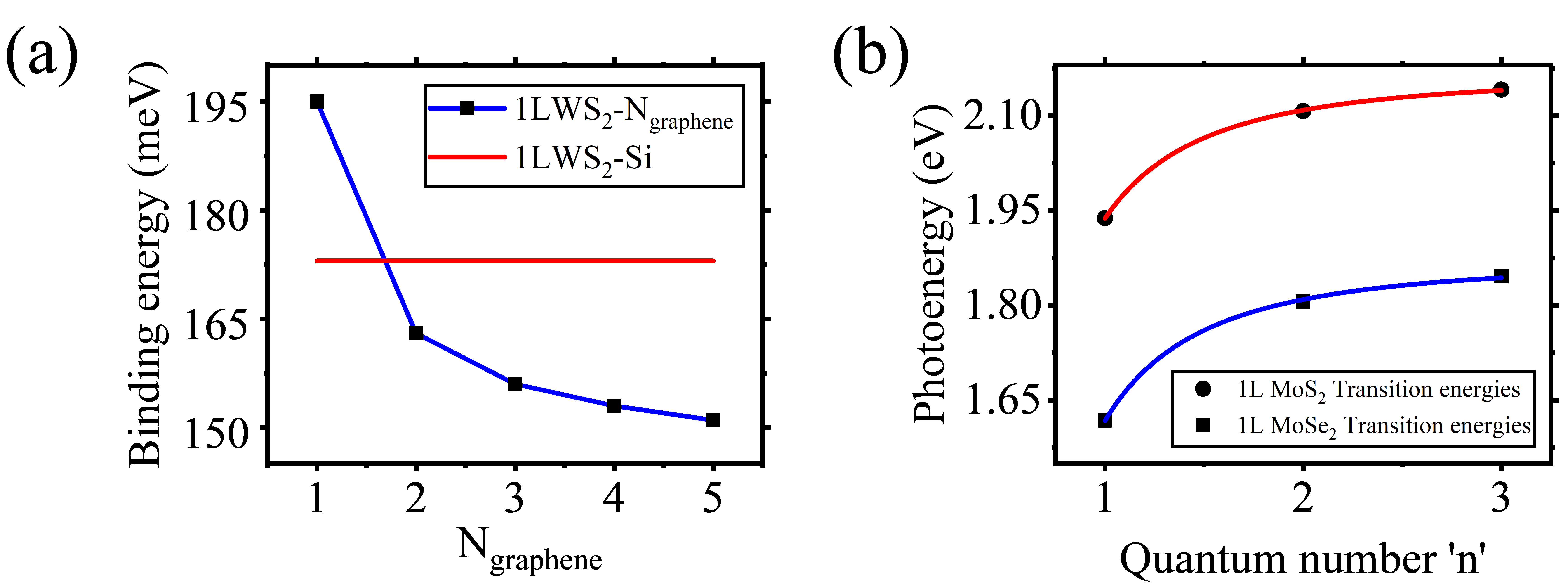}
\caption{ (a) The BE of 1L WS$_2$ as a function of the number of graphene layers as computed through the QEH model. The red horizontal line represents the BE found on the silicon substrate from the FDA fit. (b) Experimentally measured exciton RSE for monolayer MoS$_2$ and MoSe$_2$ measured by magneto-optical spectroscopy techniques as a function of `n'. The red and blue line represents the FDA fit to the data points taken from literature \cite{Goryca2019}.    }
\label{fig4}
\end{figure}

To analyze the values of BE and QBG found through the FDA method more quantitatively, we utilize the newly developed quantum electrostatic heterostructure (QEH) approach \cite{Andersen2015}. In this model applying the spatially dependent dielectric function, an electrostatic potential between the exciton constituents is obtained. In this method, including the dielectric environment is limited to van der Waals materials \cite{Andersen2015}. Thus, we employ this approach for two out of the three samples studied. Since 1L-graphene ($\epsilon_{1L-graphene}$ = 9.32 \cite{Pereira2023}) has a dielectric constant that is close to that of the silicon ($\epsilon_{Si}$ = 11.7 \cite{Dunlap1953}), we computed the BE for 1L-WS$_2$ as a function of the number of graphene layers as the substrate. The results of this computation are shown in Figure \ref{fig4} (a). We notice that the BE rapidly decreases when increasing the graphene layer numbers. Increasing the substrate graphene layer number from 1 to 2 leads to a decrease of around 30 meV in the BE, whereas transitioning from 4 to 5 graphene layers leads to only a few meV reduction. This points towards the importance of the spatial extent in the out-of-plane direction that needs to be considered for understanding the role of the screening caused by the dielectric surroundings. In Figure \ref{fig4} (a), the exciton BE corresponding to the silicon substrate is represented as a horizontal line, which intersects the BE vs graphene number of layers curve at an intermediate value (1.62) between 1 and 2 layers of graphene. Thus, the modification in BE and QBG caused by the silicon substrate is the same as that resulting from $\approx$ 2 graphene layers. The BE of 173 meV found through the FDA method on a silicon substrate is close to the reported BE of 214 meV for 1L WS$_2$ encapsulated by 2L graphene at 70 K \cite{Raja2017}. For the sample encapsulated by $\approx$ 14 nm of hBN on Si substrate, the QEH model could now be directly applied after replacing the Si substrate with 2 layers of graphene. For the top hBN encapsulation of 13.28 nm, we used a configuration consisting of 40 layers of hBN. The BE calculated using the QEH approach for this configuration of 40L hBN- 1L WS$_2$ -2L graphene is 135 meV. This agrees with the BE of 143 meV as calculated from the FDA method for the hBN encapsulated 1L WS$_2$ on silicon substrate.

\begin{table}[ht]
\centering
\begin{threeparttable}
\caption{Summary of the fractional dimensional analysis for 1L MoS$_2$ and 1L MoSe$_2$. The fractional dimension $\alpha$ and the average dielectric constant of the environment $\kappa$  are shown. Additionally, the QBG and  B.E values as calculated from the FDA fit and found experimentally in the original experiment \cite{Goryca2019} are summarized.}
\label{Table2}
\vspace*{5mm}
\begin{tabular}{|m{2.2 cm}|m{1.0 cm}|m{2 cm}|m{2 cm}|m{1.3 cm}|m{1.9 cm}|m{1.7 cm}|}
\hline \text {2D Material} &  \text {$\alpha$}  & FDA-QBG  (eV) &  FDA-BE (meV)
& $\kappa$  &  Exp-QBG (eV) &   Exp-BE (meV) \\
\hline
hBN-1L MoS$_2$-hBN & 2.96 & 2.164 $\pm$ 0.001  & 227 $\pm$ 1 & 4.45 & 2.160 & 221 \\
\hline
hBN-1L MoSe$_2$-hBN  & 2.93 & 1.870 $\pm$ 0.003  & 252 $\pm$ 3 & 4.45 & 1.874 & 231 \\
\hline
\end{tabular}
\begin{tablenotes}

\item[a]  The $\epsilon_{eff}$ value as found from the FDA fit for 1L MoS$_2$ and 1L MoSe$_2$ encapslated in hBN is 4.12 and 4.47, respectively.
\end{tablenotes}
\end{threeparttable}
\end{table}

In order to firmly establish the importance of the FDA method for 2D TMDCs other than WS$_2$, we applied it to find the QBG and BE for 1L MoS$_2$ and 1L MoSe$_2$ for which we used the available experimental data of their Rydberg series at zero magnetic field as found by magneto-optical spectroscopy in the literature \cite{Goryca2019}. The FDA fit to the Rydberg states (1s-3s) for the monolayer of MoS$_2$ and MoSe$_2$  is shown in Figure \ref{fig4} (b). The $\mu$, exciton reduced mass used is 0.275 and 0.350 for MoS$_2$ and MoSe$_2$ as calculated by performing magneto-optical spectroscopy at 91 T \cite{Goryca2019}. The details of the FDA for these 2D TMDCs are summarized in Table \ref{Table2}. The BE ratio for the 1s:2s:3s states for the MoS$_2$ and MoSe$_2$ are $1:\dfrac{1}{3.98}:\dfrac{1}{9.86}$ and $1:\dfrac{1}{3.87}:\dfrac{1}{10.95}$ respectively. These ratios are similar to the BE ratios for 3D Coulomb potential which is 1:$\dfrac{1}{4}:\dfrac{1}{9}$. Thus, the dimensionality of $\approx$ 3 as calculated through FDA matches well with the dimension found from the Rydberg series ladder. 

%In conclusion, we have successfully established the applicability of the FDA to describe the Rydberg state energies in 1L WS$_2$ and to calculate the QBG and BE in various dielectric surroundings. We demonstrate a decrease of  46\% (188 meV) in BE and a reduction of 141 meV in QBG by changing the $\kappa$ from 1.52 to 8.1. The fractional dimension calculated for the 1L WS$_2$ remained close to 2.8 even for the increase of the average dielectric of the environment by five times. The FDA may also apply to trion states in 2D semiconductors and further investigation in that direction can be a subject of future study.

In summary, our study not only establishes the robust applicability of the FDA to characterize Rydberg state energies in monolayer WS$_2$ but also unveils a profound influence of dielectric environments on excitonic properties. The achieved 46\% reduction (118 meV) in BE and a substantial 141 meV decrease in QBG by varying $\kappa$ from 1.52 to 8.1 underscore the sensitivity of excitons to their surroundings.
Remarkably, our findings transcend the specific dielectric conditions, showcasing the versatility of the FDA. The fractional dimension, consistently approximating 2.8, signifies the resilience of the monolayer WS$_2$ excitonic system to changes in the average dielectric environment, even when increased fivefold.
Furthermore, our work paves the way for potential extensions to trion states in 2D semiconductors, opening avenues for future investigations in this promising direction. The demonstrated efficacy of the FDA in unraveling excitonic behaviors in diverse scenarios positions it as a valuable tool for the broader exploration of 2D materials.
In conclusion, our results not only contribute significantly to the fundamental understanding of excitons in 2D semiconductors but also set the stage for further breakthroughs in tailoring excitonic properties for diverse applications. We anticipate that our work will inspire new avenues of research and exploration in the exciting realm of 2D materials and their unique electronic characteristics.

\begin{suppinfo}

Details of the sample fabrication, optical spectroscopy, fractional dimension calculation, optical spectroscopy results of 2D WS$_2$, atomic force microscopy of hBN-1LWS$_2$-Si sample, and dielectric screening length calculation are provided in the supporting information. 

\end{suppinfo}

\begin{acknowledgement}

We acknowledge the financial support of the Academy of Finland Flagship Programme (PREIN) (320165). C.R.F acknowledges the European Union’s Horizon 2020 research and innovation programme under the Marie Skłodowska-Curie grant agreement No. 895369.
%H.C. acknowledges the European Research Council (Starting Grant project aQUARiUM Agreement No. 802986). 

\end{acknowledgement}


\section*{Data availability}

The authors declare that the data supporting the findings of this study are available within the paper and in the supplementary material files.

\section{Competing interests}
The authors declare no competing interests.
\bibliography{bibliography}
%TC:endignore
\end{document}